\documentclass[%
aip,jap,
 amsmath,amssymb,floatfix,
 preprint,%
]{revtex4-1}
\usepackage{multirow}
\usepackage{amsmath,amssymb,amsfonts}
\usepackage{algorithmic}
\usepackage{array}
\usepackage{xcolor}
\usepackage{footnote}
\usepackage{threeparttable}
\usepackage{graphicx}
\usepackage{textcomp}
\usepackage{xcolor}
\usepackage{array}
\usepackage{mathtools}
\usepackage{siunitx}

\DeclarePairedDelimiter\abs{\lvert}{\rvert}%

\DeclareMathOperator{\sinc}{sinc}
\def\BibTeX{{\rm B\kern-.05em{\sc i\kern-.025em b}\kern-.08em
    T\kern-.1667em\lower.7ex\hbox{E}\kern-.125emX}}
\begin{document}

\title{Spin Wave Normalization Towards all Magnonic Circuits}
\author{Abdulqader Mahmoud}
\affiliation{Delft University of Technology, Department of Quantum and Computer Engineering, 2628 CD Delft, The Netherlands}

\author{Frederic Vanderveken}
\affiliation{KU Leuven, Department of Materials, SIEM, 3001 Leuven, Belgium}
\affiliation{Imec, 3001 Leuven, Belgium}

\author{Christoph Adelmann}
\affiliation{Imec, 3001 Leuven, Belgium}

\author{Florin Ciubotaru}
\affiliation{Imec, 3001 Leuven, Belgium}

\author{Sorin Cotofana}
\affiliation{Delft University of Technology, Department of Quantum and Computer Engineering, 2628 CD Delft, The Netherlands}

\author{Said Hamdioui}
\affiliation{Delft University of Technology, Department of Quantum and Computer Engineering, 2628 CD Delft, The Netherlands}

\begin{abstract}
The key enabling factor for Spin Wave (SW) technology utilization for building ultra low power circuits is the ability to energy efficiently cascade SW basic computation blocks. SW Majority gates, which constitute a universal gate set for this paradigm, operating on phase encoded data are not input output coherent in terms of SW amplitude, and as such, their cascading requires information representation conversion from SW to voltage and back, which is by no means energy effective.  In this paper, a novel  conversion free SW gate cascading scheme is proposed that achieves SW amplitude normalization by means of a directional coupler.  After introducing the normalization concept, we utilize it in the implementation of three simple circuits and, to demonstrate its  bigger scale potential, of a $2$-bit inputs SW multiplier. The proposed structures are validated by means of the Object Oriented Micromagnetic Framework (OOMMF) and GPU-accelerated Micromagnetics (MuMax3). Furthermore, we assess the normalization induced energy overhead and demonstrate that the proposed approach consumes $20$\% to $33$\% less energy  when compared with the transducers based conventional counterpart. Finally, we introduce a normalization based SW $2$-bit inputs multiplier design and compare it with functionally equivalent SW transducer based and \SI{16}{nm} CMOS designs. Our evaluation indicate that the proposed approach provided $26$\% and $6.25$x energy reductions when compared with the conventional approach and  \SI{16}{nm} CMOS counterpart, respectively, which demonstrates that our proposal is energy effective and opens the road towards the full utilization of the SW paradigm potential and the development of SW only circuits.
\end{abstract}

\maketitle

\section{Introduction}
The information technology revolution resulted in a huge amount of data that need to be processed. The processing of these data requires efficient computing platforms, which are usually implemented in CMOS technology \cite{data1,data2}. By the continuous CMOS downscaling, the performance requirements were met \cite{ITRS}. However, CMOS downscaling became more difficult due to: (i) leakage wall \cite{cmosscaling2,cmosscaling3}, (ii) reliability wall \cite{cmosscaling1}, and (iii) cost wall \cite{cmosscaling1,cmosscaling2}, which suggests that Moore's law will soon come to its end. Therefore, new technologies, such as tunnelling FETs, memristors, and spintronics \cite{survey1,survey2} are explored. A subfield of spintronics is the Spin Wave (SW) based technology \cite{survey1,survey2,ITRS}. It has three main features, which make it very promising and potentially suitable for ultra-low power consumption applications \cite{survey1,survey2,ITRS}: (i) Ultra-low power consumption because no current flows and thus no Joule heating is present, (ii) acceptable delay, (iii) scalability as SW wavelength can reach down to few nanometers at rf-frequencies. Therefore, new  design methodologies appropriate for spin-wave based technology circuits, e.g., gate cascading, which is the enabling factor towards the construction of complex SW circuits, are of great interest.

Up to date, various SW based logic gates have been proposed \cite{logic21,logic12,logic11,logic17,logic25,logic4,logic16,logic18,logic24,Magnon_transistor, logic1, logic13,logic14,logic20,logic19,logic2,logic3,logic100,logic101}. The Mach-Zehnder interferometer was used to design the first experimental SW logic gate \cite{logic21}. The same approach was used to design XNOR, NAND, and NOR gates \cite{logic12,logic11,logic17}. Also, a transmission line based three terminal device was employed to build NOT, OR, and AND gates \cite{logic25}\cite{logic4}\cite{logic16}\cite{logic18}. In addition, voltage-controlled XNOR and NAND gates were presented using a re-configurable nano-channel SW device \cite{logic24}, and two magnon transistors were embedded between the Mach-Zehnder interferometer arms to build an XOR gate \cite{Magnon_transistor}. As opposed to the previous mentioned schemes, which encode information in SW amplitude, alternative buffer, inverter, (N)AND, (N)OR, XOR and Majority gate designs  were proposed that are encoding the information in SW phase instead. \cite{logic1}. Moreover, Majority gate designs that optimize SWs transmission efficiency by decreasing their back propagation \cite{logic13,logic14,logic20}, a crossbar structure appropriate for (N)OR gate implementations \cite{logic19}, and Majority gate physical realizations \cite{logic2,logic3,logic100,logic101} were reported. 

However, the direct cascading  of two or more such logic gates within the spin wave domain is not straightforward because of the fact that they are not input-output consistent, i.e., the amplitude at the output SW originating from the input SWs interference is input data dependent, which can induce wrong results at the following gate outputs. Note that although  SW based circuits e.g., counter\cite{counter}, prime factorization \cite{prime} and multiplexer \cite{MUX}, were recently published,  all of them rely on the assumption that cascading can be performed without providing actual solutions for it. They even disregarded the issue and considered that SW gates can be directly connected, which in some cases generates wrong results as gate output SWs have input data dependent amplitude levels, or assumed that it can be achieved by forth-and-back conversions between SW and voltage domains, which is a power hungry process that may nullify the SW based computation paradigm energy efficiency promise. 

In this paper, we enable direct gate cascading within the SW domain by introducing a conversion free SW normalization approach, which opens the road towards  magnetic domain only circuit designs. The contribution of this paper can be summarized as follows:\\
\begin{itemize}
\item Enabling spin wave gate cascading through directional coupler: a properly designed directional coupler \cite{DC} is utilized to achieve logic gate SW output amplitude normalization and to pass it to the next gate.

\item Proposing and analyzing different logic gate cascading structures: Domain conversion free cascading schemes for in-line \cite{inline} and fanout enabled ladder shaped \cite{fanout} Majority gates. 

\item Building a SW based multiplier using directional coupler: We employed the cascading solution to build a $2$-bit inputs spin wave multiplier. 

\item Validating the functionality: OOMMF and MuMax3 simulations are utilized to validate all the proposed structures and evaluate their delay and energy consumption. 

\item Assessing the structures: While the proposed gate cascading solution consumes negligible amount of energy, it induces an \SI{150}{ns} delay overhead, which we reduced to \SI{20}{ns} by structure down scaling and using a material with higher average SW group velocity. In comparison with the conversion based cascading our method provides a $20$\% to $33$\% gate level energy reduction, which for the $2$-bit inputs SW multiplier results in  $26$\% and $6.25$x energy reductions when compared with the SW conventional approach and  \SI{16}{nm} CMOS counterpart, respectively.
\end{itemize}

The paper consists of eight main sections as follows. Section II discusses the basics and background of spin wave technology. Section III introduces and analyzes the gate cascading problem, Section IV explains the proposed solution, and Section V illustrates the construction of cascaded  gates and circuits. Section VI explains the simulation platform, the performed simulations, and the utilized metrics. Section VII illustrates the simulation results, provides a performance comparison when assuming the $2$-bit inputs SW multiplier as discussion vehicle, and provides inside on variability and thermal effects on SW gates functionality. Finally, Section VIII concludes the paper. 

\section{Spin Wave Basics and Background}
\label{sec:Basics of spin-wave technology}
This section provides basic inside into the spin-wave fundamentals and spin-wave based computation paradigm.

\subsection{Spin Wave Fundamentals}
\label{sec:spin-wave fundamentals}

\begin{figure}[t]
\centering
  \includegraphics[width=\linewidth]{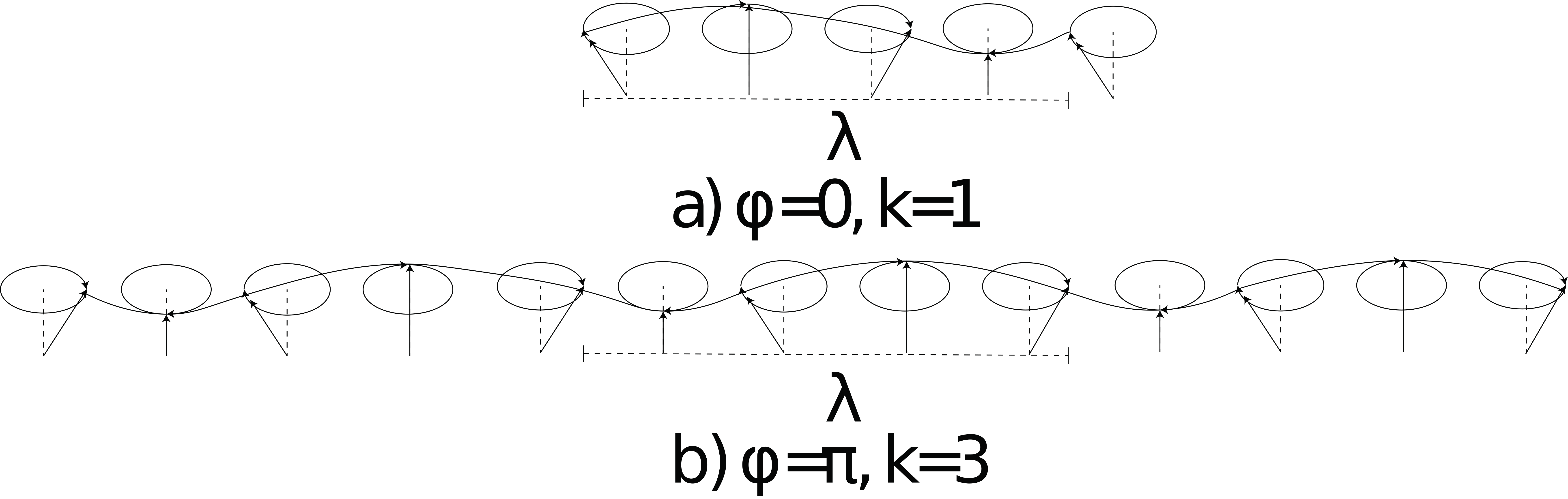}
  \caption{Spin Wave Parameters.}
  \label{fig:SW_characterstics}
\end{figure}

A spin wave is the collective excitation of the magnetization in the magnetic system \cite{Magnonic_crystals_for_data_processing}. The magnetization precessional motion can be described by using the Landau-Lifshitz-Gilbert equation \cite{LL_eq}\cite{G_eq}:

\begin{equation} \label{eq:1}
\frac{d\vec{m}}{dt} =-\abs{\gamma} \mu_0 \left (\vec{m} \times \vec{H}_{eff} \right ) + \frac{\alpha}{M_s} \left (\vec{m} \times \frac{d\vec{m}}{dt}\right ),
\end{equation}
where $\alpha$ is the damping constant, $\gamma$ the gyromagnetic ratio, $M_s$ the saturation magnetization, $\vec{m}$ the magnetization, and $H_{eff}$ the effective field. This effective field is the summation of all different field contributions that affect the magnetization. Considering the most common interactions, one obtains
\begin{equation} \label{eq:100}
H_{eff}=H_{ext}+H_{ex}+H_{demag}+H_{ani},
\end{equation}
where $H_{ext}$ is the external field, $H_{ex}$ the exchange field, $H_{demag}$ the demagnetizing field, and $H_{ani}$ the magneto-crystalline field.

Spin waves can be characterized by amplitude $A$, phase $\phi$, frequency $f$ (the time it takes for the spin to complete one round), wavelength $\lambda$ (the shortest distance between two similar spins which exhibit the same behaviour), and wavenumber $k=\frac{2 \pi}{\lambda}$ (the number of waves in one cycle, which is one full spin precision) as it can be observed in Figure \ref{fig:SW_characterstics}.

\begin{figure}[t]
\centering
  \includegraphics[width=\linewidth]{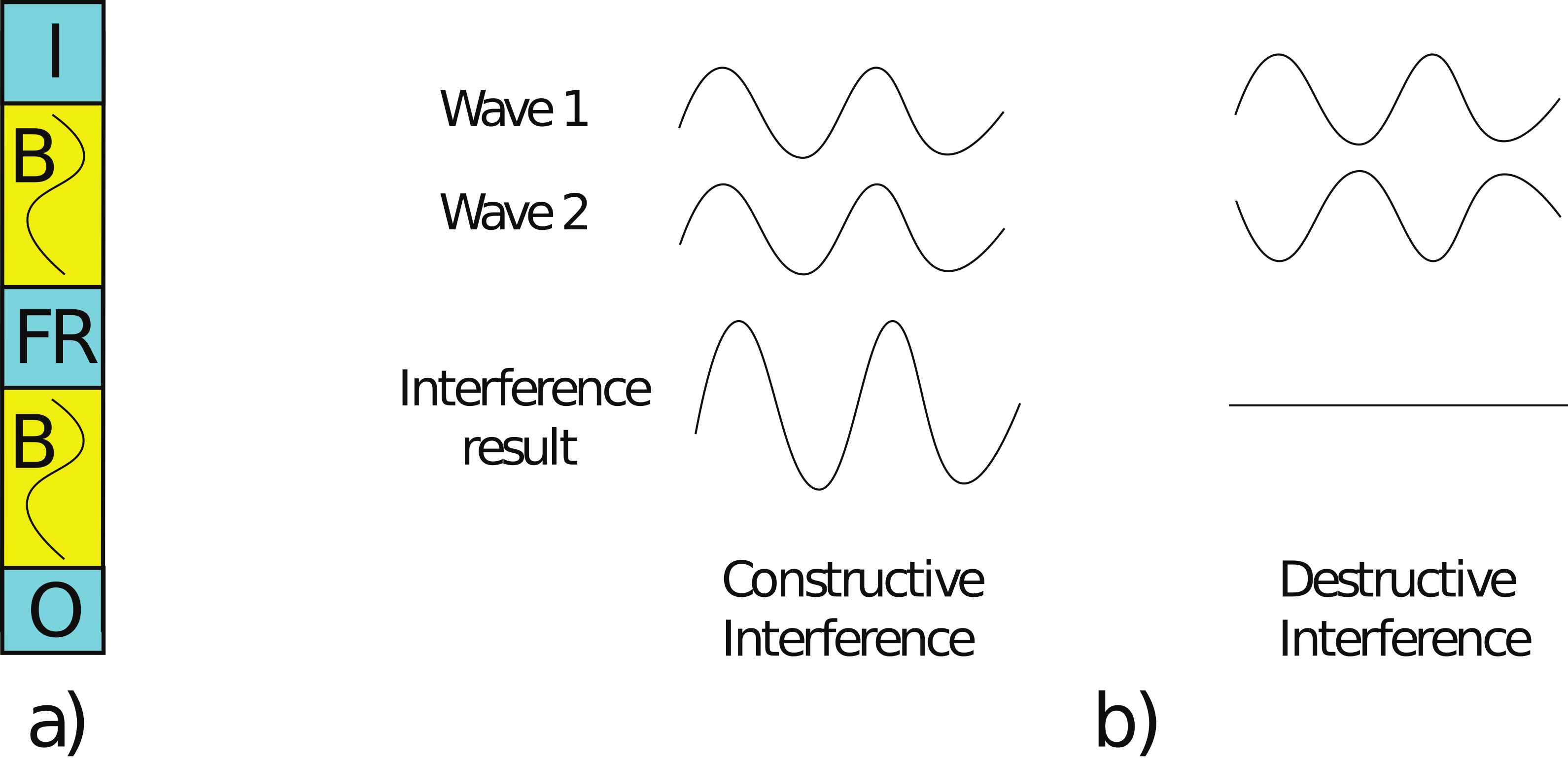}
  \caption{a) Spin Wave Device, b) Constructive and Destructive Interference.}
  \label{fig:spin_wave_device}
\end{figure} 

\subsection{Spin Wave Computing Paradigm}
\label{sec:Spin Wave Computing Paradigm}
Figure \ref{fig:spin_wave_device}a presents a spin-wave logic device. It consists of four regions: $I$, exciting stage where a spin wave is excited by, e.g., Antenna, Magneto-Electric (ME), $B$, waveguide through which the spin wave propagates, $FR$, functional region where the spin wave can be amplified, normalized, interferes with other spin waves, and $O$, the detection stage where the result is detected and converted into voltage by, e.g., Antenna, Magneto-Electric (ME) \cite{Magnonics}\cite{Magnonic_crystals_for_data_processing} \cite{logic1}. Note that SWs can be used as data carriers as during their excitation,  information can be encoded into their amplitude or phase at different  frequencies \cite{parallel_data_processing1,counter}.  In addition, SWs interference can be utilized as underlying principle behind SW  computing strategies that do not follow the well establish Boolean algebra paradigm. To get inside into this operation principle we make use of  the interference of two SWs as discussion vehicle. Their interference is constructive if they are in phase $\Delta \phi=0$, and destructive if they are out of phase $\Delta \phi=\pi$, as depicted in Figure \ref{fig:spin_wave_device}b. Subsequently, assuming that logic $0/1$ is represented by a spin wave with phase $0$/$\pi$ and more than two waves coexist in the same waveguide, the majority principle governs their interference. Assuming for example that $3$ SWs are reaching the $FR$ and that at most one of them has a phase of $\pi$, then the resulting SW has a $0$ phase and of $\pi$ otherwise, which mimics the $3$-input Majority gate behaviour. Note that while in the SW domain $3$-input Majority can be evaluated with one device only its CMOS implementation requires $18$ transistors \cite{logic1,logic9},  which clearly indicates that SW based implementation are potentially speaking more compact and energy effective than CMOS counterparts. 

\begin{figure}[t]
%\vspace{-0.5cm}
\centering
  \includegraphics[width=\linewidth]{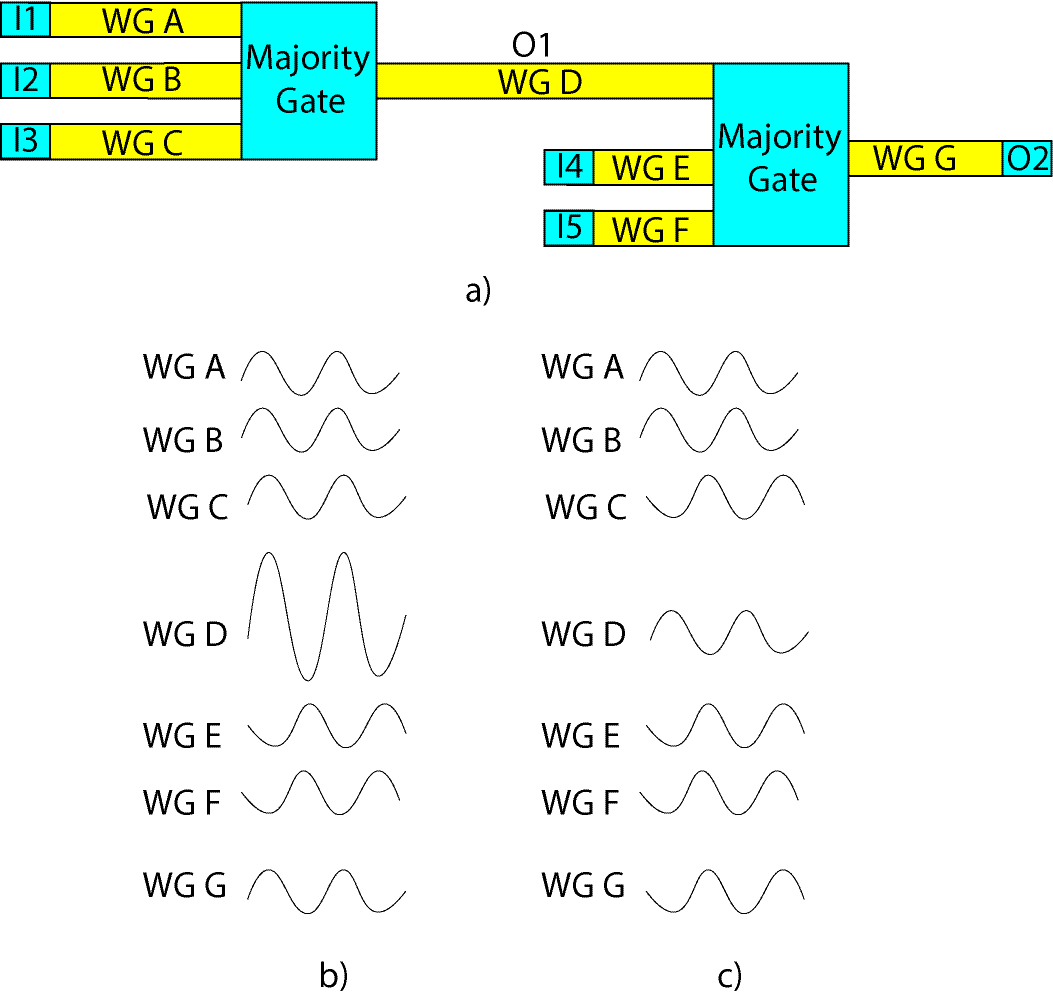}
  \caption{a) Cascaded MAJ3 Gates, Spin Wave Waveform Analysis at b) $I_1I_2I_3I_4I_5$=$00011$, c) $I_1I_2I_3I_4I_5$=$00111$.}
  \label{fig:cascading}
\end{figure}

\section{Spin Wave Gate Cascading Challenge}

To evaluate complex Boolean functions, one needs to be able to interconnect spin wave gates to form the required circuit. However, directly cascading Majority or any other type of SW gates may produce wrong results. To clarify this issue let as assume the situation in  Figure \ref{fig:cascading}a where a $3$-input Majority (MAJ3) gate output is connected to one of the inputs of another MAJ3 gate. All input SW are excited with the same amplitude $A$, frequency $f$, and a $0$ phase corresponds to logic $0$  and a $\pi$ phase to logic $1$. Given that MAJ3  operation is governed by SW interference both amplitude and phase of the SW gate inputs contribute to the output SW parameters. While from the point of view of an individual gate the output value is solely determined by the output SW phase this is not any longer the case when that output is utilized as input for a followup gates. Figure \ref{fig:cascading}b and c present the SW interferences within the circuit when $I_1I_2I_3I_4I_5=00011$ and $I_1I_2I_3I_4I_5=00111$, respectively.  As one can observe in Figure \ref{fig:cascading}b the excited spin waves at $I_1$, $I_2$, and $I_3$ interfere constructively and produce on WG D a spin wave with the same phase as $I_1$ $I_2$, and $I_3$, but with a $3A$ amplitude (strong majority). Subsequently, WG D SW  interacts with $I_4$ and $I_5$ SWs in the second MAJ3 gate, which produces an output SW with amplitude $A$ and phase $0$, which is wrong given that $MAJ3(0,1,1) = 1$. This wrong results is induced by the fact that the MAJ3 gate can properly operates on equal amplitude SWs, which is not the case for  $I_1I_2I_3I_4I_5=00011$. Figure  \ref{fig:cascading}c present the situation for $I_1I_2I_3I_4I_5=00111$ case in which the first MAJ3 produces an $A$ amplitude and phase $0$ SW (weak majority) and the second gate produces the correct result as expected. Thus, cascading MAJ3 may induce wrong output results when the driving gate produces a strong majority $0$ or $1$ output. 

\begin{figure}[t]
\centering
  \includegraphics[width=\linewidth]{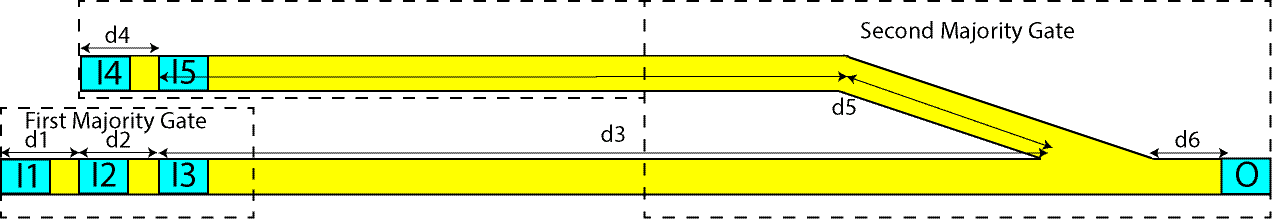}
  \caption{Cascaded In-Line MAJ3 Gates.}
  \label{fig:cascading1}
\end{figure} 

\begin{figure}[t]
\centering
  \includegraphics[width=\linewidth]{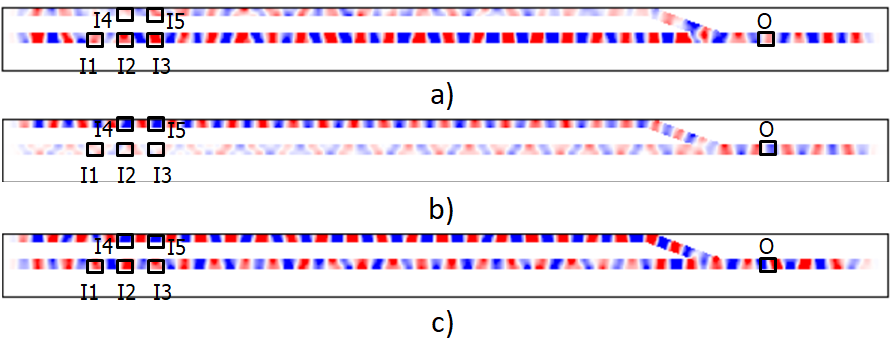}
  \caption{Cascaded In-Line MAJ3 Gates Simulation Results.}
  \label{fig:results1}
\end{figure} 
To clarify things even more, we build the structure depicted in Figure \ref{fig:cascading1} that corresponds to two cascaded MAJ3 gates and evaluated its behaviour by means of  OOMMF simulations.  Figure \ref{fig:results1} presents the OOMMF results when  the parameters mentioned in Section V are utilized. Three different cases were tested $I_1I_2I_3I_4I_5=00000$, $I_1I_2I_3I_4I_5=00111$, and $I_1I_2I_3I_4I_5=00011$. In the Figure, red represents logic $0$, and blue  logic $1$. As it can be observed from the figure, $I_1I_2I_3I_4I_5=00000$ results in an output $O=0$, while $I_1I_2I_3I_4I_5=00111$ resulted in an output $O=1$. However, in the case of $I_1I_2I_3I_4I_5=00011$, the output is between logic $0$ and logic $1$ as a result of the strong $0$ generated by the first MAJ3 gate (SW with $3A$ amplitude). Thus as the theoretical analysis also suggested wrong results are generated, which call for the MAJ3 gate augmentation with an amplitude normalizer able to enable SW gates cascading and, by implication, circuit design in the spin wave domain.

\section{Proposed SW Gate Cascading Solution}
\label{sec:Spin Wave Gate Cascading}
This section first introduces the proposed gate cascading concept and its operation principles.  Thereafter, it demonstrates its capability to circumvent the problem presented in the previous section and illustrated in Figure \ref{fig:cascading}.

\subsection{Proposed SW Gate Cascading Concept}
\label{sec:Directional Coupler}
\begin{figure}
\centering
  \includegraphics[width=0.5\linewidth]{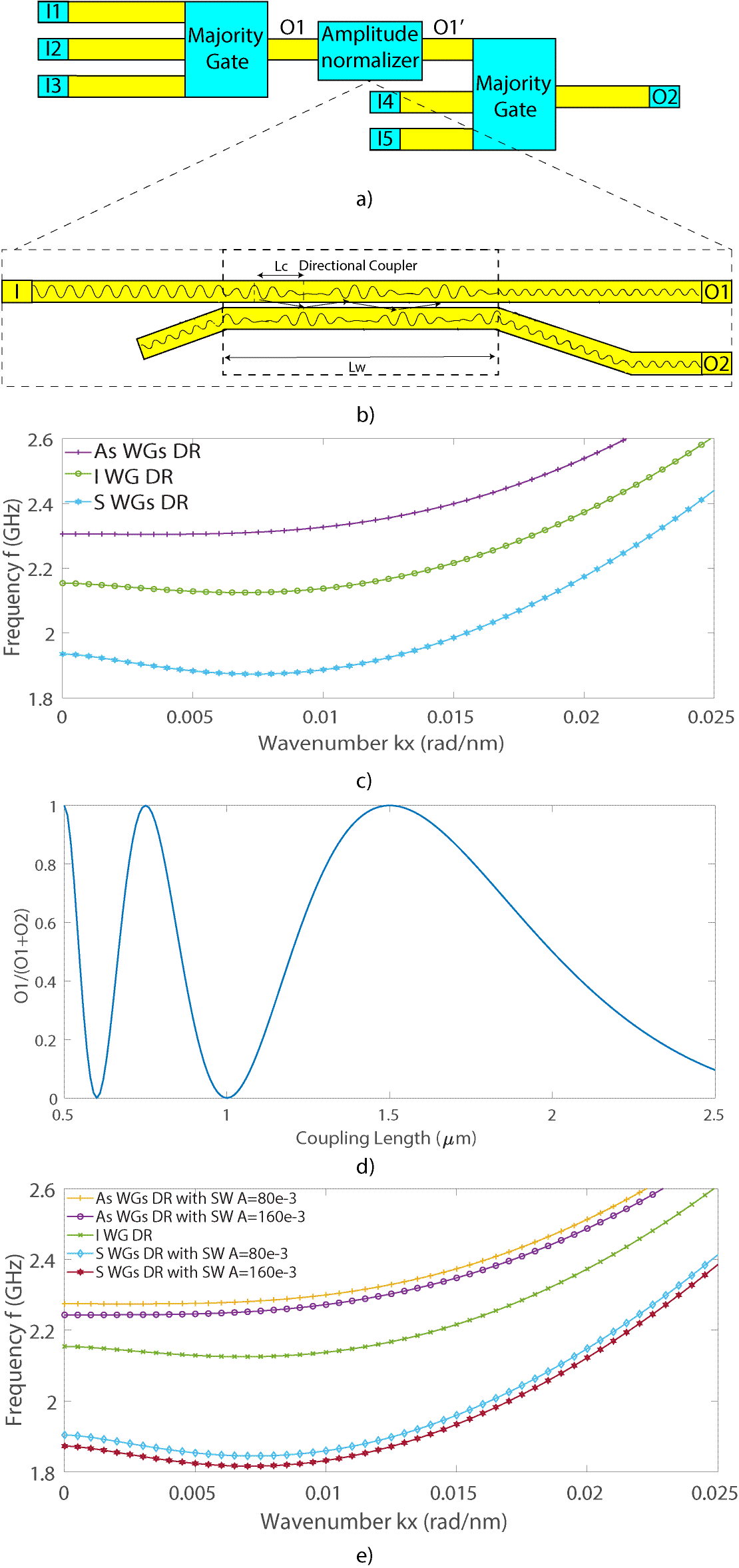}
  \caption{a) Proposed Spin Wave Gate Cascading Solution, b) Directional Coupler, c) Dispersion Relation (DR) of Isolated (I), Symmetric (S) and Asymmetric (As) Spin Wave Waveguide (WG) Modes at the Linear Region, d) Energy Transmission Ratio between Coupled Waveguides with $L_w$=3 $\mu$ m  , e) Dispersion Relation of Single, Symmetric and Asymmetric Spin Wave Waveguide modes at the Non-linear region (with Frequency Shift Effect).}
  \label{fig:DC}
\end{figure}

The proposed gate cascading solution relies on the placement of a spin wave amplitude normalizer between the cascaded  Majority gates. The normalizer is a properly designed directional coupler \cite {DC} able to adjust the driving Majority gate output SW amplitude to $A$ in case of strong majority ($3A$) or to leave it unchanged for weak majority cases  before passing it to the next Majority gate as presented in Figure \ref{fig:DC}a. This behaviour is achieved by making use of the nonlinear properties of high amplitude SWs, which cause a shift in the dispersion relation,  which at its turn induces a  wavelength shift. When placing two waveguides close to each other they are said to be dipolarly coupled and form a directional coupler as presented in Figure \ref{fig:DC}b, which enables a wavelength dependent energy transfer between the two waveguides. Thus, by properly controlling this energy transfer via the nonlinear characteristics, the spin wave amplitude can be normalised to the desired value, i.e., $A$ in our case.

The equations describing the dispersion relations and energy transfer of the normaliser element are given in the following. A detailed derivation of the equations can be found in \cite{DC6,DC7,DC,DC1}.  When two waveguides are placed close to each other, two spin wave modes exist. One mode has a symmetric profile over both waveguides whereas the other has an antisymmetric profile over the two waveguides. The dispersion relation of both modes is given by
\begin{equation} \label{eq:2}
f_o(k_x) = \frac{1}{2\pi}\sqrt{\Omega^{yy}\Omega^{zz}}
\end{equation}
and
\begin{equation} \label{eq:3}
f_{s,as}(k_x) = \frac{1}{2\pi}\sqrt{(\Omega^{yy}\pm \omega_MF_{kx}^{yy}(d))(\Omega^{zz}\pm \omega_MF_{kx}^{yy}(d))},
\end{equation}

\noindent where $f_o(k_x)$ is the SW dispersion relation in a single waveguide, $f_{s,as}(k_x)$ the symmetric and anti-symmetric dispersion relations for spin waves in coupled waveguides, $\Omega^{ii}=\omega_H + \omega_M(\lambda_{ex}^2k_x^2+F_{kx}^{ii}(0))$, $i=y,z$, $\omega_H =\gamma B_{ext}$, $\omega_M =\gamma \mu_o M_s$, $M_s$ the magnetic saturation, $\gamma$ the gyromagnetic ratio, $\mu_o$ the vacuum permeability, $\lambda_{ex}=2A_{ex}/\mu_oM_s^2$, $A_{ex}$ the exchange constant, $d=w+\delta$ the distance between the two waveguides centres, $w$ the waveguides width, $\delta$ the gap between the two waveguides, and $\overset{\wedge}F_{kx}$ the tensor that describes the dynamical magneto-dipolar interaction (introduced in \cite{DC6,DC7,DC,DC1})
\begin{equation} \label{eq:4}
F_{kx}^{yy}(d)=\frac{1}{2\pi}\int{\frac{\abs{\sigma}^2k_y^2}{\tilde{w}k^2}(1-\frac{1-e^{-kh}}{kh})e^{ik_yd}dk_y},
\end{equation}
\begin{equation} \label{eq:5}
F_{kx}^{zz}(d)=\frac{1}{2\pi}\int{\frac{\abs{\sigma}^2}{\tilde{w}}\frac{1-e^{-kh}}{kh}e^{ik_yd}dk_y},
\end{equation}
where $k=\sqrt{k_x^2+k_y^2}$, $h$ the material thickness, $\sigma$ the Fourier transform of the spin wave profile across the width of the waveguide, and $\tilde{w}$ the mode profile normalized constant. When the spins are fully unpinned at the waveguide edges,   $\tilde{w}$ equals the real waveguide width and $\sigma=w \sinc(k_yw/2)$.

When a spin wave is excited at frequencies higher than the anti-symmetric mode minimum frequency, two spin wave modes are excited at the same time. One symmetric mode with wavenumber $k_s$ and antisymmetric mode with wavenumber $k_{as}$. As a result of the interference between them, the overall spin wave energy resonantly transfers from one waveguide to the other after SW's propagation over a particular distance $L_c$ as depicted in Figure \ref{fig:DC}b \cite{DC,DC1,DC2,DC3,DC4}. This distance $L_c$ is called coupling length, and depends on different parameters such as SW wavelength, applied magnetic field, space between waveguides,  waveguide geometrical size, and SW amplitude \cite{DC}. The coupling length is given by \cite{DC,DC1}
\begin{equation} \label{eq:6}
L_c = \frac{\pi}{\abs{k_s-k_{as}}}.
\end{equation}

The distribution of SW energy over the two waveguides at the end of the normaliser depends on the coupling length $L_c$ and the length of the coupled waveguides $L_w$. 
The proportion of energy in the first waveguide after a distance $L_w$ is given by  \cite{DC}
\begin{equation} \label{eq:7}
\frac{O_1}{O_1+O_2} = \cos^2\left(\frac{\pi L_w}{2L_c}\right),
\end{equation}
where $O_1$ and $O_2$ are the output energies of the first and second waveguide, 

as also graphically visualised in Figure \ref{fig:DC}d.

As long as the SW amplitude is low, the nonlinear effects are limited. However, as the spin wave amplitude increases, the nonlinearity affects the spin wave dispersion relation, and causes a frequency shift. This dispersion relation corresponding to nonlinear spin waves is given by
\begin{equation} \label{eq:8}
f^{(nl)}_{s,as} = f^{(0)}_{s,as}(k_x)+T_{kx}\abs{a_{kx}}^2,
\end{equation}
where $a_{kx}$ is the spin wave amplitude and $T_{kx}$ the spin wave nonlinear frequency shift coefficient, which can be calculated by \cite{DC5,DC8,DC,DC1}

\begin{equation} \label{eq:9}
T_{kx}=\frac{w_H-A_{kx}+\frac{B_{kx}^2}{2\omega_o^2}(\omega_M(4\lambda^2k_x^2+F_{2kx}^{xx}(0))+3\omega_H)}{2\pi}
\end{equation}
with 
\begin{equation}
A_{kx}=\omega_H+\frac{\omega_H}{2}(2\lambda_{ex}^2k_x^2+F_{kx}^{yy}(0)+F_{kx}^{zz}(0)) \,,
\end{equation}

\begin{equation} B_{kx}=\frac{\omega_M}{2}(F_{kx}^{yy}(0)-F_{kx}^{zz}(0)) \,,
\end{equation}
and 
\begin{equation}
F_{2kx}^{xx}(d)=\frac{1}{2\pi}\int{\frac{\abs{\sigma}^24k_x^2}{\tilde{w}k^2}(1-\frac{1-e^{-kh}}{kh})e^{ik_yd}dk_y}
\end{equation}

with $k=\sqrt{4k_x^2+k_y^2}$.

 This is also graphically presented in Figure \ref{fig:DC}e \cite{DC6,DC}. Note that the parameters we utilize for determining these dispersion relations are summarized in Table I.

The nonlinear frequency shift also affects the distribution of the energies over the two waveguides as indicated by 
\begin{equation} \label{eq:10}
\frac{O_1}{O_1+O_2} = \cos^2\left(\frac{\pi L_w}{2L}-\frac{\pi L_w}{2L^2}\frac{\partial L}{\partial f}T_{kx}\abs{a_{kx}}^2\right).
\end{equation}

As it is clear from Equation (\ref{eq:10}), the nonlinear effects of the spin waves strongly influence the power distribution over the two waveguides. Hence, the directional coupler exhibits high sensitivity to spin wave amplitude changes. As a result, if a strong coupling and high sensitivity to the spin wave amplitude change are required, the directional coupler must be long and the gap between the two directional couplers must be small. 
For example, if $0\%$,  $50\%$, and $100\%$ of the input spin wave energy should transfer to the second waveguide when its amplitude is $2A$, $3A$, and $4A$, respectively, $L_w$ should be equal to \SI{3}{\mu m}, the distance between the coupled waveguide (DW) \SI{10}{nm}, Yttrium Iron Garnet (YIG) waveguide thickness  \SI{30}{nm} and width  \SI{100}{nm}, wavelength  \SI{340}{nm}, and frequency  \SI{2.282}{GHz} \cite{DC1}. These values are material depend, thus they change when another material is utilized \cite{DC1}. 

Note that such a directional coupler can be utilized as frequency multiplexer and others \cite{DC}. However, in this paper, we concentrate on its utilization as amplitude normalizer to enable gate cascading within spin wave domain.

\subsection{DC based SW Gate Cascading Implementation}

\begin{figure}[t]
%\vspace{-0.5cm}
\centering
  \includegraphics[width=\linewidth]{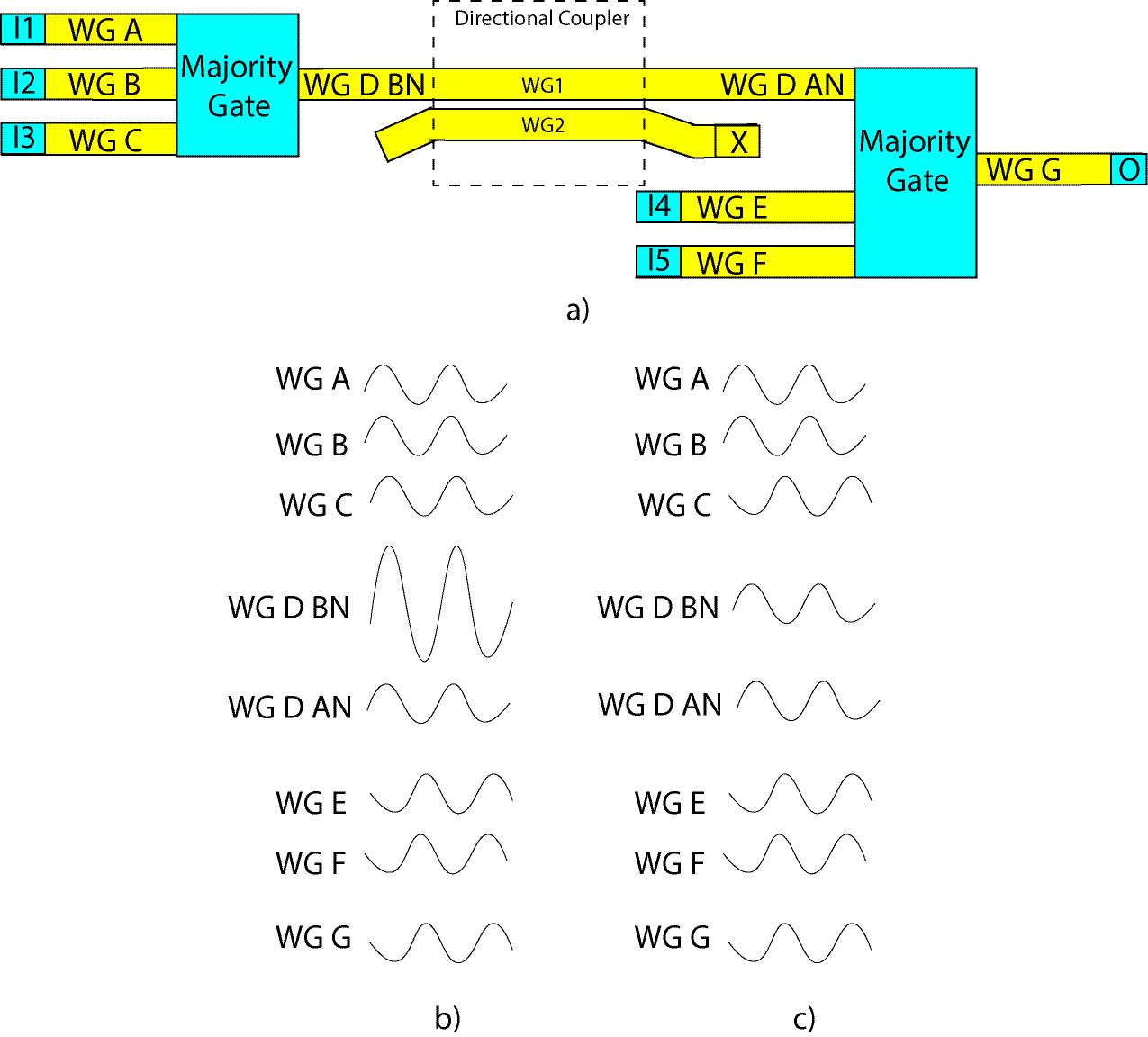}
  \caption{(a) Proposed Gates Cascading Solution. Spin Wave Waveform Analysis (b) $I_1I_2I_3I_4I_5$=$00011$, (c) $I_1I_2I_3I_4I_5$=$00111$.}
  \label{fig:structure2}
\end{figure}

Figure \ref{fig:structure2}a revisit the situation in Figure \ref{fig:cascading}a and augments the waveguide connecting the two majority gates with a directional coupler as amplitude normalizer.
The spin waves excited at $I_1$, $I_2$, $I_3$ interfere constructively or destructively depending on their phases and the output of  the first MAJ3 gate is normalized or not on case it signals a strong or a weak majority by the directional coupler. If the output SW  amplitude is greater than a predefined threshold, in our case the inputs amplitude value $A$, then it is normalized to $A$ while preserving the SW phase. Otherwise, no normalization occurs and only a tinny portion of the SW power is transfered to the second waveguide due to the coupling effect. The two input combinations we previously utilized explain the gate cascading issue, i.e., $I_1I_2I_3I_4I_5$=$00011$ and $I_1I_2I_3I_4I_5$=$00111$, are revisited to demonstrate that the directional coupler enables proper gate cascading. Assuming that all input spin waves are excited with the same amplitude $A$ and frequency ones  excited at $I_1$, $I_2$, and $I_3$ interfere constructively in the first case resulting in a spin wave with $0$ phase and $3A$ amplitude as depicted by WG D BN in Figure \ref{fig:structure2}b. Given that SW amplitude is greater than $A$ it is normalized by the directional coupler to $A$ producing WG D AN in Figure \ref{fig:structure2}b. At the second majority gate WG E and WG F  interfere constructively which result destructively interfere with WG D AN. As a result of the overall interference process the output SW corresponds to a logic $1$ as it should. In the other case, $I_1$ SW constructively interferes  with  $I_2$ SW which result    destructively interferes with $I_3$ SW resulting in  a spin wave with $0$ phase and amplitude $A$ in WG D BN. Since the amplitude equals to the threshold, no normalization occurs and the WG D AN spin wave approximately equals  WG D BN SW as depicted in Figure \ref{fig:structure2}c. Then the spin wave excited at $I_4$ and $I_5$ interfere constructively with each other and destructively with spin wave in WG D AN, which result in a $\pi$ phase and amplitude $A$ SW, i.e., a logic $1$ as expected.

\section{Building cascaded SW gates and circuits}

In order to validate our proposal and demonstrate its potential towards building spin wave circuits, we design three complex gates that make use of it. To cover the most common situations encountered in logic circuit implementations we selected three different structures for demonstration purpose, as follows:  (i) Single output MAJ3 gate and (ii)  Fully/Partially cascadable dual output MAJ3 gates. Note that the introduced approach is scalable and can be applied to SW gates with more outputs but such designs are beyond the goal of this manuscript. Additionally, in order to asses the cascading approach potential at circuit level we instantiate a $2$-bit inputs spin wave multiplier  presented in Figure \ref{fig:multiplier}, which spin wave domain only design is not possible without the proposed approach.

\begin{figure}[t]
%\vspace{-0.5cm}
\centering
  \includegraphics[width=\linewidth]{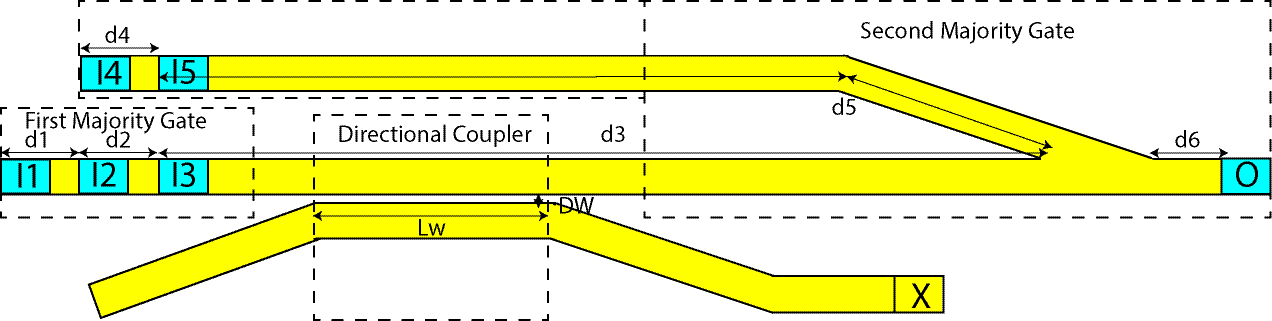}
  \caption{In-Line MAJ3 Cascaded Gates.}
  \label{fig:structure3}
\end{figure}

\subsection{Cascaded In-Line MAJ3 Gates}
The structure in Figure \ref{fig:structure2}a provides a generic gate cascading solution containing multiple bent regions, which are not SW propagation "friendly". To minimize them, we implemented the two in-line majority cascaded gates  compound with one bent region as depicted in Figure \ref{fig:structure3}. 
Note that the normalized output of the first Majority gate acts as the third input  of he second Majority gate.  

To guarantee proper results, the structure dimensions must be fulfil certain constraints as follows. If SWs should  constructively interfere  when they have the same phase and destructively otherwise,  $d_1=d_2=\ldots=d_5= n \times \lambda$, where $n = 0, 1, 2, 3, \ldots$. If the opposite behaviour is desired, i.e., SWs constructively interfere if they are out of phase and destructively  otherwise, $d_1=d_2=\ldots=d_5= (n+\frac{1}{2}) \times \lambda$.

The output of the first Majority gate must be normalized to the amplitude of the second Majority gate inputs. Assuming that all input SWs have an amplitude of $A$ the output of the first Majority gate must be normalized to $A$ in case it reports a strong majority result, i.e., a $3A$ amplitude SW.  Therefore, if the output amplitude is $A$ no normalization is required, whereas if the output amplitude is $3A$ a normalization is performed such that $66$\% of the spin wave power moves into the second waveguide towards $X$ and only $33$\% of it passes to the second Majority gate.  To obtain this bahaviour, the directional coupler is designed by making use of Equations (\ref{eq:2})-(\ref{eq:10}) while taking into consideration different parameters including applied magnetic field, spaces between waveguides, dimension of the waveguides, static magnetization orientation, and spin wave wavelength, frequency, and amplitude. 

The output position must be determined accurately to obtain the desired results, i.e.,  MAJ3 and inverted MAJ3 are obtained when   $d_{6}= n \times \lambda$ and $d_{6}= (n+\frac{1}{2}) \times \lambda$, respectively. Moreover, depending on a predefined phase,  the output value can be phase detected, i.e.,  $\Delta \phi=0$ represents logic $0$ and $\Delta \phi=\pi$ logic $1$. By following the same line of reasoning as in Section IV.B one can easily check the correct behaviour of the two in-line cascaded gates, which is also demonstrated by the simulation results presented in Section VII Figure \ref{fig:results2}.

\begin{figure}[t]
%\vspace{-0.5cm}
\centering
  \includegraphics[width=0.5\linewidth]{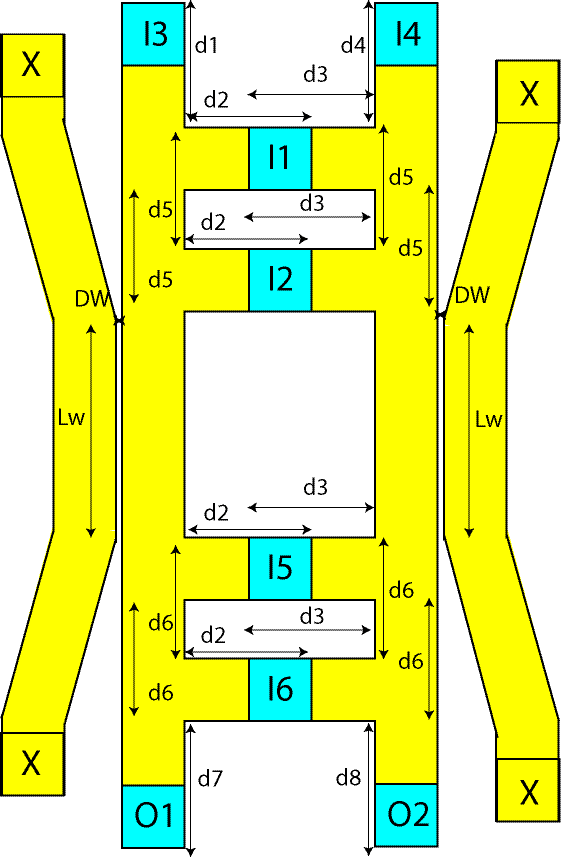}
  \caption{Fully Cascaded Ladder MAJ3 Gates.}
  \label{fig:structure4}
\end{figure}

\subsection{Fully Cascaded Ladder MAJ3 Gates}

As the efficient implementation of real life circuits requires gates with fanout capabilities a fanout of $2$ ladder shaped MAJ3 gate has been introduce in \cite{fanout}. Before discussing the augmentation of such a gate with directional couplers we briefly discuss its operation principle. 

The upper part of the structure presented in Figure \ref{fig:structure4} constitutes  a MAJ3 gate that is able to parallelly evaluate $MAJ(I_1,I_2,I_3)$ and $MAJ(I_1,I_2,I_4)$, thus if $I_3=I_4$ the two values are equal and the gate exhibits a fanout of $2$.  As discussed in  \cite{fanout} the waveguide topology and dimensions are determined in such a way that the input SWs can properly interfere and generate the correct output values, according with the Majority function true table, and the SW present in the left/right arm before the directional coupler carries the $MAJ(I_1,I_2,I_3)$/$MAJ(I_1,I_2,I_4)$ value. Simply speaking, the MAJ3 gate operates as follows: (i) At $I_1$, $I_2$, $I_3$, and $I_4$, SWs are excited with suitable phase, i.e., phase $0$ for logic $0$ and phase $\pi$ for logic $1$, (ii) Excited SWs propagate through the horizontal and vertical waveguides, (iii) At the "meeting" points, they interfere constructively or destructively depending on their phases, and (iv) Finally, the resultant SWs propagate downwards  through the left and right arms. 
 
\begin{figure}[t]
%\vspace{-0.5cm}
\centering
  \includegraphics[width=0.4\linewidth]{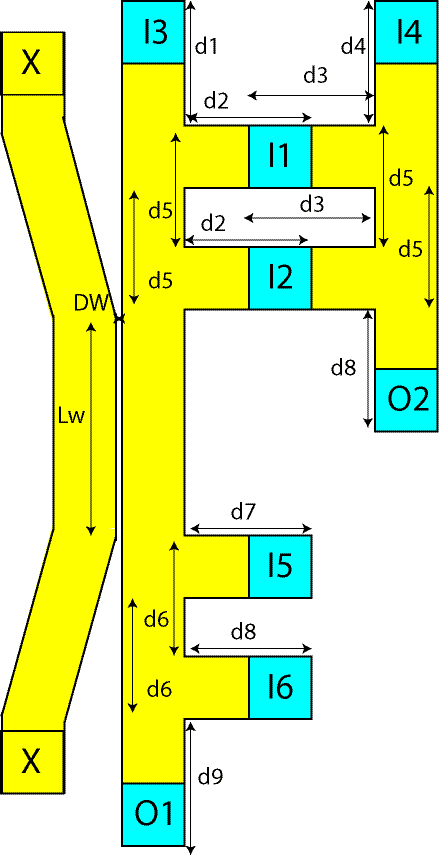}
  \caption{Partially Cascaded Ladder MAJ3 Gates.}
  \label{fig:structure5}
\end{figure}

To make the FO2 MAJ3 gate outputs directly connectable as inputs to following SW gates they have to be normalized by means of $2$ directional couplers as presented in Figure \ref{fig:structure4}. The circuit in the Figure operates as follows: (i) At $I_1$, $I_2$, $I_3$, $I_4$, $I_5$,and $I_6$, SWs are excited with suitable phase, (ii) The excited spin waves propagate horizontally and vertically and at the intersection point, they interfere constructively or destructively depending on the excited SWs phases in both arms, (iii) The resulted spin waves from the first Majority gate propagate toward the couplers to be normalized, (iv) The normalized SWs propagate downward to interfere with the spin waves excited at $I_5$ and $I_6$, and  (v) Finally, the resulted SWs propagate toward $O_1$ and $O_2$ such that $O_1=MAJ(MAJ(I_1,I_2,I_3),I_5,I_6)$ and $O_2=MAJ(MAJ(I_1,I_2,I_4),I_4,I_6)$ and that $I_3$=$I_4$. Note that in case $I_3=I_4$ the two outputs are equal, thus the gate compound exhibits a fanout of $2$, but when $I_3 \ne I_4$ the circuit evaluates two different functions that benefit circuit complexity.  

To guaranty correct behaviour the input SWs must have the same amplitude and wavelength $\lambda$, which, to simplify the interference pattern, must be greater than the waveguide width $w$. The structure dimension $d_{i}, i= 1,2,\ldots,6$ must be determined in terms of $\lambda$. For instance, if SWs have to constructively interfere when they have the same phase and destructively interfere when they are out of phase, $d_1,d_2,\ldots,d_6$ must be equal with $n \lambda$, where $n=1,2,3,...$. However, if the other way around is desired, i.e., SWs with the same phase should interfere destructively and constructively when they are out of phase, $d_1,d_2,\ldots,d_6$ must be equal with $(n+\frac{1}{2}) \lambda$, where $n=1,2,3,...$.  Additionally, the outputs can be captured at $O_1$ and $O_2$ located at $d_7$ and $d_8$ from the last interference point, which should be $n\lambda$ or $(n+\frac{1}{2})\lambda$ if the non-inverted or inverted output is desired, respectively. Note that the couplers which are needed to normalize the outputs of the first Majority gates are designed in same way as described in the previous section.

\subsection{Partially Cascaded Ladder MAJ3 Gates}

In this situation the FO2 MAJ3 gate is providing input to one follow up MAJ3 gate while its second output constitutes a circuit primary output, i.e., it is read out by a SW detection cell. Consequently, only one  directional coupler is required as depicted in Figure \ref{fig:structure5}, while the operation principle and the design steps are the same as for the previously discussed structures.

\subsection{$2$-bit Inputs Spin Wave Multiplier }
\begin{figure}[t]
\centering
  \includegraphics[width=\linewidth]{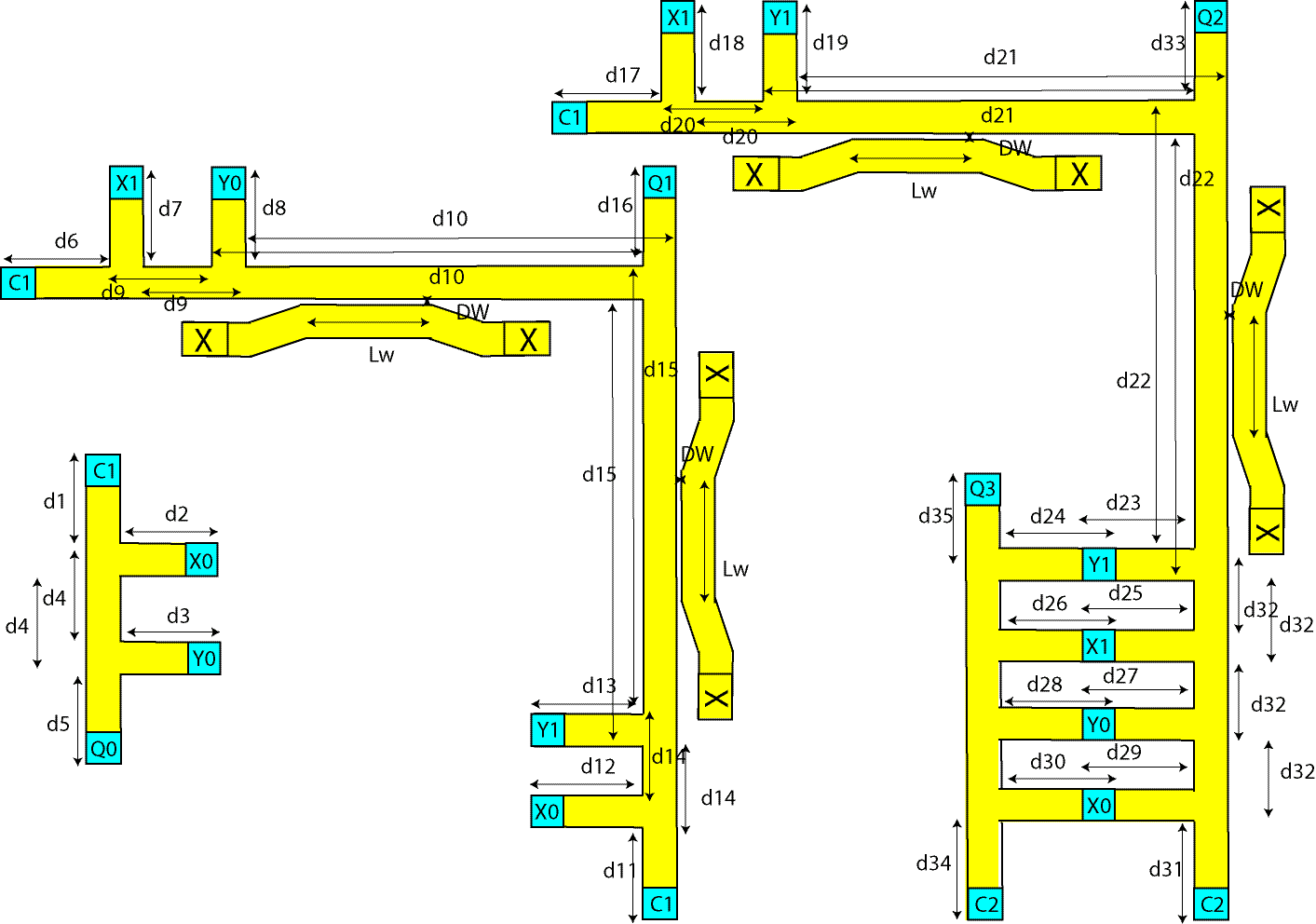}
  \caption{$2$-bit Inputs Spin Wave Multiplier.}
  \label{fig:multiplier}
\end{figure}

Figure \ref{fig:multiplier} presents a $2$-bit inputs SW multiplier that makes use of the proposed normaliser. The multiplier inputs are the operands $X = (X_1, X_0)$ and $Y = (Y_1,Y_0)$ and the control signals $C_1$ and $C_2$. The structure requires $18$ excitation cells and generates a $4$-bit output  $Q=(Q_0, Q_1, Q_2, Q_3)$. Following the multiplication algorithm  $Q_0 = AND(X_0, Y_0)$ and $Q_1 = XOR(AND(X_1, Y_0),AND(X_0, Y_1))$. $2$ directional couplers are needed to normalize  AND gates outputs to enable their cascading to the XOR gate. Further, $Q_2 = XOR(AND(X_0, Y_0),AND(X_0, Y_0,X_1, Y_1))$, and again $2$ directional couplers are required. Finally, $Q_3=AND(X_0, Y_0, X_1, Y_1)$. 

As previously discussed, the distances depend of the chosen SW wavelength and must be accurately determined, i.e.,  $d_i=n \lambda$, where $i \in \{1, 2, \ldots, 35\}$,  $n=0, 1, 2, \ldots$ and  $n \neq \{5,16, 33,35\}$ as the required interference has to interfere constructively if the SWs have the same phase, and destructively if they are out of phase $\Delta \phi=\pi$. %as it presented in Figure \ref{fig:multiplier}.

Moreover, as the circuit includes AND and XOR gates, phased based detection, briefly explained in Section V.A, is required for $Q_0$ and $Q_3$ and threshold based detection for  $Q_1$ and $Q_2$. The threshold based detection relies on comparing  the spin wave amplitude with a given value in order to discriminate between the two logic values, i.e., greater than the threshold corresponds to logic $1$ and lower to logic $0$. 
To ensure correct output detection $d_5$ and $d_{35}$ must be $n \lambda$ to read the non-inverted output. In contrast, $Q_1$ and $Q_2$  should be located as near as possible to the interference point to minimize SW amplitude attenuation.

\section{Simulation Setup}
\label{sec:Simulation Setup and Experiments}
In the following lines, the simulation platform, the utilized parameters, and the performed simulations and performance evaluation  metrics are described.
\subsection{Simulation Platform}
\label{subsec:Simulation Setup}
We make use of Object Oriented Micro Magnetic Framework (OOMMF) \cite{OOMMF} and MuMax3 \cite{mumax} to validate the correct functionality of the proposed normalization solution and gate structures. In the simulations, blue represents a logic $1$ and red a  logic $0$. \\
The  parameters provided to the micromagnetic software are presented in Table \ref{table:1} \cite{DC}. The dimension of the structures is equal to a spin wave wavelength multiple. Therefore, dimension of the structure in Figure \ref{fig:structure3} are  $d_1$=$d_2$=$d_4$=\SI{340}{nm}, $d_3$=\SI{3.74}{\mu m}, $d_5$=\SI{4.08}{\mu m}, and $d_6$=\SI{340}{nm}, whereas the dimension of the structure in Figure \ref{fig:structure4} and \ref{fig:structure5} are $d_1$=$d_2$=$d_3$=$d_4$=$d_5$=$d_6$=$d_7$=$d_8$= \SI{340}{nm} and $d_1$=$d_2$=$d_3$=$d_4$=$d_5$=$d_6$=$d_7$=$d_8$=$d_9$=\SI{340}{nm}.  Moreover, as further discussed  in the simulation results subsection, when making use of a YIG wave guide the directional coupler induced delay is $150$ ns, which can be  decreased by scaling down the structure or by utilizing  another material with higher spin wave group velocity. In this work, $Fe_{60}Co_{20}B_{20}$ was utilized as waveguide material with Perpendicular Magnetic Anisotropy (PMA). The material parameters are: magnetic saturation $M_s$=$1.1 \times 10^6$\SI{}{A/m}, exchange stiffness $A_{ex}$=\SI{18.5}{pJ/m}, damping constant $\alpha=2\times 10^{-4}$, and perpendicular anisotropy constant $k_{ani}=8.3177 \times 10^5$J/m$^3$ \cite{parameters}. The  waveguide with is \SI{30}{nm} and its thickness \SI{1}{nm}. SWs are excited at a frequency of \SI{15}{GHz} and have a wavelength of \SI{100}{nm}. In addition, as the  waveguide length should be equal to a wavelength multiple we have chosen it to be $5$ times the wavelength, i.e., \SI{500}{nm}, to decrease mutual effects of gate arms and directional couplers on each others. By making use of  Equations (\ref{eq:2})-(\ref{eq:10}) we determined the directional coupler dimensions as $L_w$=\SI{2.55}{\mu m} and $DW$=\SI{8}{nm}.

\begin{table}[t]
\caption{Simulation Parameters}
\label{table:1}
\centering
  \begin{tabular}{|c|c|}
    \hline
    Parameters & Values \\
    \hline
    Magnetic saturation $M_s$ & $1.4 \times 10^5$\SI{}{A/m} \\
    \hline
    Damping constant $\alpha$ & $0.0002$ \\
    \hline
    Waveguide thickness $t$ & \SI{30}{nm} \\
    \hline
    Exchange stiffness $A_{ex}$ & \SI{3.5}{pJ/m} \\
    \hline
    $L_w$ & \SI{3}{\mu m} \\
    \hline
    $DW$ & \SI{8}{nm} \\
    \hline
    $\lambda$ & \SI{340}{nm} \\
    \hline
    Frequency $f$ & \SI{2.282}{GHz} \\
    \hline
  \end{tabular}
\end{table}
\subsection{Performed Simulation and Evaluation Metrics}
\label{subsec:Experimental Performed}
We performed simulations on the $4$ structures introduced in Section \ref{sec:Simulation Results and Discussion}. 

Delay, power, and energy consumption and delay are metrics of interest to evaluate the gate cascading structures and the multiplier. The energy and delay of transducers are based on the estimation in \cite{Excitation_table_ref16} and the SW delay through waveguides was estimated directly from OOMMF and MuMax3 simulation results. The following assumptions are made: i) The excitation and detection cells are ME cell, i.e., $C_{ME}$=\SI{1}{fF}, $V_{ME}$=\SI{119}{mV}, Energy=$k \times C_{ME} \times V_{ME}^2$ (where $k$ is the number of excitation cells), and \SI{0.42}{ns} ME cell switching delay \cite{Excitation_table_ref16}, ii) SW consumes tiny energy in the waveguide and directional coupler when compared to the energy consumed by the transducers, and iii) SWs are excited by means of pulse signals. We note that due to the early stage development of the SW technology, these assumptions might not be accurate and the assumed values may change in the close future. %However, their further investigations is part of future work when the technology is mature enough to be investigated.

%==================================
%==================================
\section{Simulation Results and Discussion}
%===================================
%====================================
\label{sec:Simulation Results and Discussion}
In this section simulation results for the gate cascading structures and the spin wave multiplier are presented and commented upon. In addition,  delay, power, and energy overhead are assessed and compared with domain conversion and $16$ nm CMOS based functionally equivalent counterpart designs. Finally, variability and thermal effects are discussed.
\subsection{MAJ3 Gate Cascading}
\label{secSimulation Results}

\begin{figure}[t]
\centering
  \includegraphics[width=\linewidth]{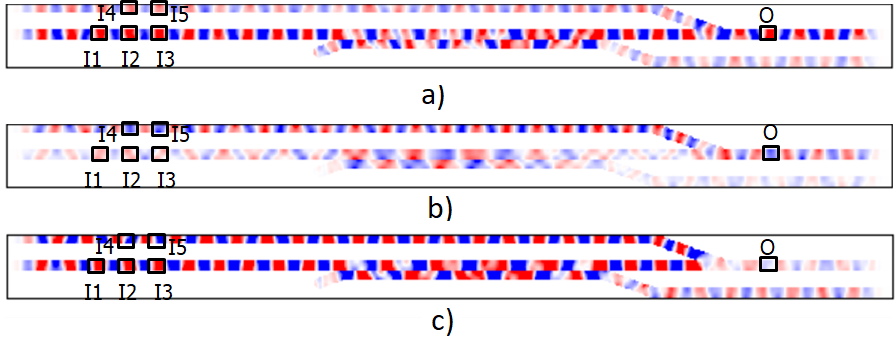}
  \caption{Cascaded In-line MAJ3 Gates: (a) $I_1I_2I_3I_4I_5 = 00000$, (b) $I_1I_2I_3I_4I_5 = 00111$, and (c) $I_1I_2I_3I_4I_5 = 00011$.}
  \label{fig:results2}
\end{figure}

\subsubsection* {\textbf{In-Line MAJ3 Gates}}

Figure \ref{fig:results2} (a), (b), and (c) presents the simulation results of the two MAJ3 inline cascaded gates (see Figure \ref{fig:structure3} for the input patterns $I_1I_2I_3I_4I_5 = 00000$,  $I_1I_2I_3I_4I_5 = 00111$, and  $I_1I_2I_3I_4I_5=00011$, respectively). By inspecting the Figures, it is clear the output results are as expected, i.e.,  the output corresponding to $I_1I_2I_3I_4I_5 = 00000$ is logic $0$ because all inputs are logic $0$ and logic $1$ in the other cases because two inputs of the second Majority gate are logic $1$ and one input is logic $0$, due to the proper amplitude correction induced by the directional coupler. 

\subsubsection*{\textbf{Fully Cascaded Ladder MAJ3 Gates}} 
\begin{figure}[t]
\centering
  \includegraphics[width=0.5\linewidth]{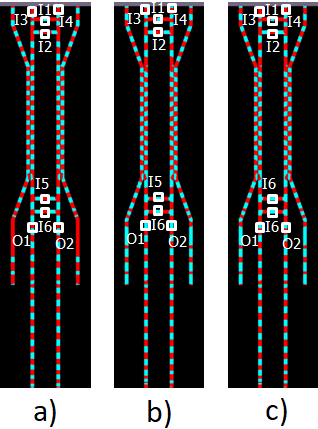}
  \caption{Fully Cascaded Ladder MAJ3 Gates: (a) $I_1I_2I_3I_4I_5 = 00000$, (b) $I_1I_2I_3I_4I_5 = 00111$, and (c) $I_1I_2I_3I_4I_5 = 00011$.}
  \label{fig:results4}
\end{figure}

Figure \ref{fig:results4} (a), (b), and (c) presents the MuMax3 simulation results for the structure in Figure \ref{fig:structure4} corresponding to $2$ fully cascaded ladder MAJ3 gates for the input combinations $I_1I_2I_3I_4I_5I_6 = 000000$, $I_1I_2I_3I_4I_5I_6 = 001111$ , and   $I_1I_2I_3I_4I_5I_6 = 000011$, respectively. It is clear from the Figure that the outputs $O_1$ and $O_2$ are correct, i.e., $O_1=O_2=0$ when  $I_1I_2I_3I_4I_5I_6 = 00000$ because all circuit inputs are logic $0$, while $O_1=O_2=1$ when  $I_1I_2I_3I_4I_5I_6 = 001111$ and  $I_1I_2I_3I_4I_5I_6 = 000011$ because two inputs of the second MAJ3 gate are logic $1$ and the other logic $0$, which demonstrates the correct behaviour of the circuit. 

\subsubsection*{\textbf{Partially Cascaded Ladder MAJ3 Gates}} 
\begin{figure}[t]
\centering
  \includegraphics[width=0.5\linewidth]{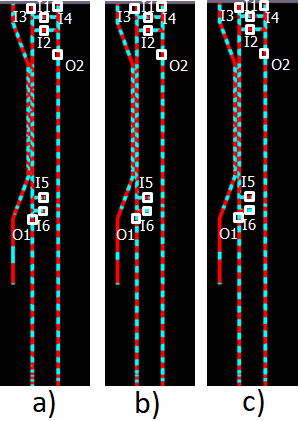}
  \caption{Partially Cascaded Ladder MAJ3 Gates: (a) $I_1I_2I_3I_4I_5 = 00000$, (b) $I_1I_2I_3I_4I_5 = 00111$, and (c) $I_1I_2I_3I_4I_5 = 00011$.}
  \label{fig:results6}
\end{figure}

Figure \ref{fig:results6} (a), (b), and (c) presents the MuMax3 simulation results for the structure in Figure \ref{fig:structure5} corresponding to the partial cascadation of $2$  ladder MAJ3 gates for the input combinations  $I_1I_2I_3I_4I_5I_6 = 000000$,  $I_1I_2I_3I_4I_5I_6 = 001111$, and $I_1I_2I_3I_4I_5I_6 = 000011$, respectively. By inspecting the figures, it is clear that all cases $O_1$ assumes the correct value (for $I_1I_2I_3I_4I_5I_6 = 00000$ is logic $0$ because all inputs are logic $0$ and logic $1$ in the other cases because two inputs of the second MAJ3 gate are logic $1$ and the third one logic $0$. On the other arm, which is not cascaded with the second  MAJ3 gate, $O_2$ is not normalized and correct results are obtained $O_2$ (logic $0$ in all cases as $I_5$ and $I_6$ do not affect its behaviour. %, when $I_1I_2I_3I_4I_5I_6 = 000000$, $I_1I_2I_3I_4I_5I_6 = 001111$, and $I_1I_2I_3I_4I_5I_6 = 000011$ because all inputs are logic $0$.\\

\subsubsection*{\textbf{$2$-bit Inputs Spin Wave Multiplier}} 
\begin{figure}[t]
\centering
  \includegraphics[width=\linewidth]{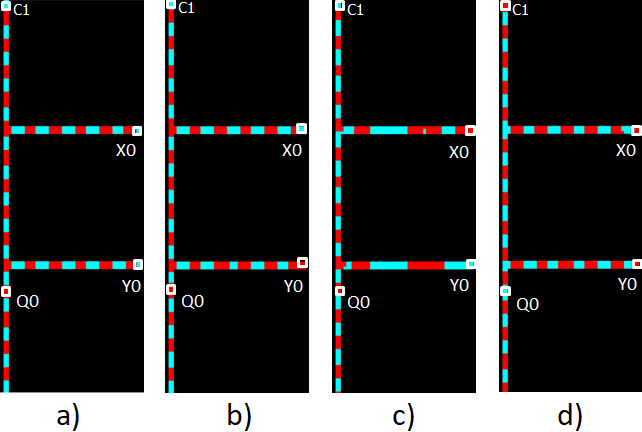}
  \caption{$Q_0$ Output Simulation (a) $X_0Y_0 = 00$, (b) $X_0Y_0 = 01$, (c) $X_0Y_0 = 10$, and (d) $X_0Y_0 = 11$.}  \label{fig:results8}
\end{figure} 

The $2$-bit inputs spin wave multiplier in Figure \ref{fig:multiplier} is validated by MuMax3 using the same parameters as for the $30$nm width $Fe_{60}Co_{20}B_{20}$ waveguide in the previous subsection. 

Figure \ref{fig:results8} presents the first output $Q_0$ simulation results. Note that $Q_0 = AND(X_0,Y_0) = MAJ(0,X_0,Y_0)$ thus $C_1$ in Figure \ref{fig:multiplier} should be asserted to $0$. 

Inspecting Figure \ref{fig:results8} reveals $Q_0$'s correct behaviour.  Note that $Q_0$ is placed at $d5 = 510$nm ($n=5$). 

\begin{table}[t]
%\vspace{-0.2cm}
\caption{Normalized Second and Third Spin Wave Multiplier Outputs.}
\label{table:q1}
\centering
  \begin{tabular}{|c|c|c|c|c|c|c|}
    \hline
   \multicolumn{4}{|c|}{Cases} & $Q_1$ & $Q_2$  \\ \hline
    $X1$ & $Y1$ & $X0$ & $Y0$ &  &   \\
    \hline
    $0$ & $0$ & $0$ & $0$ & $0.03$ & $0.06$ \\
    \hline
    $0$ & $0$ & $0$ & $1$ & $0.08$ & $0.03$  \\
    \hline
    $0$ & $0$ & $1$ & $0$ & $0.22$ & $0.016$  \\
    \hline
    $0$ & $0$ & $1$ & $1$ & $0.15$ & $0.04$ \\
    \hline
     $0$ & $1$ & $0$ & $0$ & $0.38$ & $0.17$ \\
    \hline
    $0$ & $1$ & $0$ & $1$ & $0.03$ & $0.3$  \\
    \hline
    $0$ & $1$ & $1$ & $0$ & $0.46$ & $0.09$  \\
    \hline
    $0$ & $1$ & $1$ & $1$ & $0.74$ & $0.09$  \\
    \hline
    $1$ & $0$ & $0$ & $0$ & $0.32$ & $0.3$ \\
    \hline
     $1$ & $0$ & $0$ & $1$ & $1$ & $0.16$ \\
    \hline
    $1$ & $0$ & $1$ & $0$ & $0.1$ & $0.006$ \\
    \hline
    $1$ & $0$ & $1$ & $1$ & $0.54$ & $0.0003$ \\
    \hline
    $1$ & $1$ & $0$ & $0$ & $0.002$ & $1$  \\
    \hline
    $1$ & $1$ & $0$ & $1$ & $0.52$ & $0.7$  \\
    \hline
    $1$ & $1$ & $1$ & $0$ & $0.52$ & $0.33$  \\
    \hline
    $1$ & $1$ & $1$ & $1$ & $0.22$ & $0.2$ \\
    \hline
  \end{tabular}
\end{table}

\begin{figure}[t]
\centering
  \includegraphics[width=\linewidth]{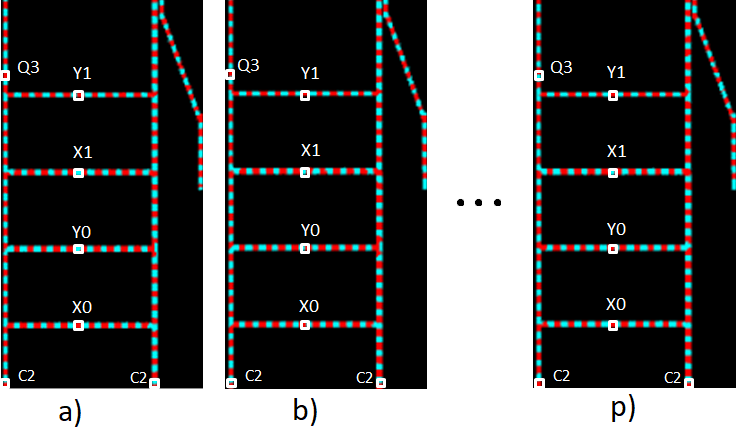}
  \caption{Fourth Spin Wave Multiplier Output (a) $X1Y1X0Y0 = 0000$, (b) $X1Y1X0Y0 = 0001$, and (p) $X1Y1X0Y0 = 1111$.}
  \label{fig:results9}
\end{figure} 

As $Q_1$ and $Q_2$ are computed as XOR functions threshold detection is required to determine their values and as such Table \ref{table:q1} presents $Q_1$ and $Q_2 $ normalized spin wave magnetization for different inputs combinations $X_0Y_0X_1Y_1 = 0000$, $X_0Y_0X_1Y_1 = 0001$, \ldots, and $X_0Y_0X_1Y_1 = 1111$.  Note that to achieve proper circuit functionality $C_2$ SW amplitude has to be higher that the one of input SW by a factor of $2.25$, which is the required value the realization of the $4$-input AND over the input bits. 
%{\bf Check this again please. 
In order to implement the threshold detection, an appropriate threshold is determined for each output, i.e.,  the normalized threshold for $Q_1$ is $0.42$, and for $Q_2$ is $0.315$. As presented in the table, as the inputs combinations $X_0Y_0X_1Y_1 = 0000$, $X_0Y_0X_1Y_1 = 0001$, $X_0Y_0X_1Y_1 = 0010$, $X_0Y_0X_1Y_1 = 0011$, $X_0Y_0X_1Y_1 = 0100$, $X_0Y_0X_1Y_1 = 0101$, $X_0Y_0X_1Y_1 = 1000$, $X_0Y_0X_1Y_1 = 1010$, $X_0Y_0X_1Y_1 = 1100$, and $X_0Y_0X_1Y_1 = 1111$ results in output magnetization less than the threshold, thus $Q_1 = 0$, and $Q_1 = 1$ for $X_0Y_0X_1Y_1 = 0110$, $X_0Y_0X_1Y_1 = 0111$, $X_0Y_0X_1Y_1 = 1110$, $X_0Y_0X_1Y_1 = 1001$, $X_0Y_0X_1Y_1 = 1011$, and $X_0Y_0X_1Y_1 = 1101$ because these input combinations result in output spin wave amplitudes larger than the threshold. Also, as the inputs combinations $X_0Y_0X_1Y_1 = 0011$, $X_0Y_0X_1Y_1 = 0111$, and $X_0Y_0X_1Y_1 = 1011$ result in output magnetization greater than the threshold, thus $Q_2 = 1$, and $Q_2 = 0$ for the rest cases. Note that the normalized thresholds average for $Q_1$ and $Q_2$ are obtained by averaging the normalized magnetization for $Q_1$ and $Q_2$ between inputs 0001 and 1001 for $Q_1$ and inputs 1011 and 0101 for $Q_2$.

Figure \ref{fig:results9} presents the forth output $Q_3$ simulation results for $X_0Y_0X_1Y_1 = 0000$, $X_0Y_0X_1Y_1 = 0001$, \ldots, and $X_0Y_0X_1Y_1 = 1111$. As it can be observed in the Figure $Q_3$, which is $AND(X_0,Y_0,X_1,Y_1)$, is  correctly evaluated. 

\subsection{Performance Evaluation} 

Whereas normalization based cascading doesn't consume a noticeable amount of energy, in comparison with transducers based counterpart (no ME cells for domain conversion are required and the electrons are not moving but just spin and affect each other by the dipolar coupling effect), it induces a significant delay overhead. To estimate the delay, i.e.,  the maximum time it takes for the SW outputs to become available for further processing, we make use of the numerical simulation results and for all YIG waveguides based  considered  structures we computed a coupler induced delay of \SI{150}{ns}. 

\begin{figure}[t]
\centering
  \includegraphics[width=\linewidth]{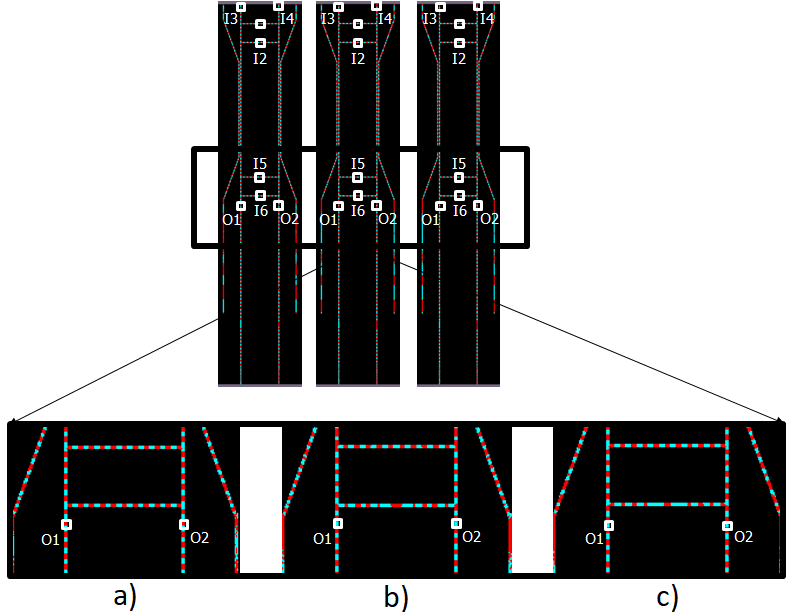}
  \caption{Scaled Down Fully Cascaded MAJ3 Gates at (a) $I_1I_2I_3I_4I_5 = 00000$,  (b) $I_1I_2I_3I_4I_5 = 00111$, and (c) $I_1I_2I_3I_4I_5 = 00011$.}
  \label{fig:results7}
\end{figure}

Although this delay overhead is rather large, it can be decreased by structure downscaling and by relying on alternative materials with higher SW group velocity and/or  other coupling effects than dipolar, which is slow by its nature.
To get an indication on the scaling effect, we validated by means of MuMax3 simulations the cascading of FO2 MAJ3  gates constructed with $Fe_{60}Co_{20}B_{20}$ waveguides of \SI{30}{nm} width. Simulation results for  $I_1I_2I_3I_4I_5I_6 = 000000$,  $I_1I_2I_3I_4I_5I_6 = 001111$, and $I_1I_2I_3I_4I_5I_6 = 000011$ are presented in Figure \ref{fig:results7} and one can easily check that the output values are correct. 
Remarkable is the fact that scaling and material change diminished the delay overhead from \SI{150}{ns} to \SI{20}{ns}, which  indicates that the overhead can potentially be further decreased towards the \SI{}{ps} range.

In order to evaluate the practical implications of our proposal we evaluate coupler and conversion based cascading and compare them in terms of delay, power, and energy consumption. The conversion based circuits are obtained by replacing each directional coupler in Figures \ref{fig:structure3}, \ref{fig:structure4}, and \ref{fig:structure5} with two transducers able to convert SW to charge domain and back. Given that the assumptions in Section VI.B the following conjectures are utilized in the evaluations: (i) Transducers (MEs) are the main contributor to the circuit power consumption as  the power consumption related to SWs propagation trough waveguide and directional coupler are insignificant, (ii) SW propagation delay in the waveguide is neglected, (iii) ME transduce power consumption and delay are \SI{34.3}{\mu W} and \SI{0.42}{ps}, respectively \cite{Excitation_table_ref16}, and iv) SWs are excited by means of pulse signals. 

For delay calculations we identify the critical path length through each considered structure.  As this spans over $2$ ME cells and one directional coupler, and $4$ ME cells for coupler and conversion based designs, respectively, the delay sums up to \SI{20.84}{ns} and \SI{1.68}{ns}, respectively.

As SW propagation, interference, and normalization are assumed to happen at zero power costs the power consumed by each design is determined by the number of ME cells it includes. Given that conversion based designs require $8$, $12$, and $10$ ME cells, the power sums up to  \SI{274.4}{\mu W}, \SI{411.6}{\mu W}, and \SI{343}{\mu W} for the in-line, ladder fully, and ladder partially cascaded structures, respectively. On the other hand coupler based structures require $6$, $8$, and $8$ ME cells which results in  \SI{205}{\mu W}, \SI{274.4}{\mu W}, and \SI{274.4}{\mu W} for the in-line, ladder fully, and ladder partially cascaded structures, respectively. 

Finally, the energy consumption can be derived as the power-delay product. We note however that due to pulse operation paradigm ME activation follows the domino behaviour. Thus each of them is active for a short period of time necessary for its output SW creation, i.e., the ME cell delay of \SI{0.42}{ns} under current assumptions, and idle for the rest of the calculation. This means that regardless of the overall circuit delay the energy should be evaluated as the product of power consumption and ME cell delay. By following this procedure the energy consumed by conversion based the in-line, ladder fully, and ladder partially cascaded structures is derived as \SI{115.2}{aJ}, \SI{172.8}{aJ}, and \SI{144}{aJ}, respectively, and  \SI{86.4}{aJ}, \SI{115.2}{aJ}, and \SI{115.2}{aJ} for the coupling based counterparts.

\begin{table}[t]
\caption{Comparison with cascading based conversion}
\label{table:11}
\centering
\begin{threeparttable}
  \begin{tabular}{|>{\centering}m{4em}|>{\centering}m{2.5em}|>{\centering}m{2.5em}|>{\centering}m{3em}|>{\centering}m{2.5em}|>{\centering}m{2.5em}|>{\centering}m{3em}|}
    \hline
     & \multicolumn{3}{c|}{Conversion cascading}  & \multicolumn{3}{c|}{Coupler cascading} \tabularnewline \hline
%     Technology &  SW & SW & SW & SW & SW & SW \tabularnewline  \hline
     Structure &  IL & LFC & LPC & IL & LFC & LPC \tabularnewline
    \hline
    Power (\SI{}{\mu W}) &  $274.4$ &  $411.6$ & $343$ & $205$ & $274.4$ & $274.4$ \tabularnewline
    \hline
     Delay (\SI{}{ns}) &  $1.68$ &  $1.68$ & $1.68$ & $20.84$ & $20.84$ & $20.84$ \tabularnewline
    \hline
    Energy\tnote{1} (\SI{}{aJ}) &  $115.2$ &  $172.8$ & $144$ & $86.4$ & $115.2$ & $115.2$ \tabularnewline
    \hline
  \end{tabular}
  \begin{tablenotes}
  \item[1]  Due to pulse mode operation each ME is active for the time necessary for its output SW creation and idle for the rest of the calculation. Thus, regardless of the overall circuit delay, the energy is evaluated as the product of power consumption and the ME cell delay (\SI{0.42}{ns}).
  \end{tablenotes}
  \end{threeparttable}
\end{table}

Table \ref{table:11} presents the comparison of the coupler and conversion based implementations in terms of power, delay, and energy consumption. In the Table IL, LFC, and LPC, stand for In-Line, Ladder Fully Cascaded, Ladder Partially Cascaded structures, respectively. As expected, the coupler based approach provides a power reduction of $25$\%, $33$\%, and $20$\% for in-line, ladder fully, and ladder partially cascaded circuits, respectively. Moreover, given that pulse SW operation is utilized  the directional coupler delay overhead is not negatively affecting the energy consumption and the same savings are obtained in therms of energy. Note that the coupler based cascading may become more delay effective by further scaling down the structure, and the utilization of other materials and/or faster coupling effects.

To get more inside into the potential implications of our proposal we compare the proposed $2$-bit inputs multiplier with SW conversion based and \SI{16}{nm} CMOS implementation counterparts. 

The CMOS implementation requires $6$ AND and $2$ XOR gates  and its delay and energy consumption are estimated based on the figures reported in \cite{16nmCMOS}. The SW implementation for coupler based cascading is the one described in Figure \ref{fig:multiplier} and the implementation for the conversion based cascading is designed by replacing each directional coupler with two transducers to convert SW to charge domain and back. The assumptions and calculation methodology utilized for $2$ MAJ3 circuits comparison are in place.

Table \ref{table:6} presents the comparison in terms of energy and delay between the $3$ considered $2$-bit inputs multiplier implementations. As it can be observed in the Table, spin wave implementations are more energy efficient than the \SI{16}{nm} CMOS counterpart, i.e., $6.25\times$ and $4.65\times$  less energy for coupler and conversion based cascading, respectively. Moreover, the proposed solution consumes $31$\% less energy than the approach relying on forth and back conversion between spin wave and charge domains, while having a $12.5\times$ larger delay. Although the proposed solution is much slower, its main strong point is the ultra low energy consumption enabled by the directional coupler utilization.  As previously mentioned the delay can be reduced by scaling and the utilization of other materials and/or faster coupling effect, thus we are still far from reaching the  ultimate energy consumption reduction horizon.

\begin{table}[t]
\caption{$2$-bit Input Multiplier Performance.}
\label{table:6}
\centering
  \begin{tabular}{|>{\centering}m{5em}|>{\centering}m{5em}|>{\centering}m{6em}|>{\centering}m{5em}|}
    \hline
    Technology &  \SI{16}{nm} CMOS &  SW &  SW \tabularnewline
    \hline
    Topology  &  \SI{16}{nm} CMOS &  Conversion Cascading &  Coupler Cascading \tabularnewline
    \hline
   %  Implemented function &  2x2 bit multiplier & 2x2 bit multiplier & 2x2 bit multiplier \tabularnewline \hline
     Energy (\SI{}{fJ}) &  $2$ &  $0.43$ &  $0.32$ \tabularnewline
    \hline
     Delay (\SI{}{ns}) &  $0.1$ &  $1.68$ &  $21$ \tabularnewline
    \hline
  \end{tabular}
 \vspace{-0.2cm}
\end{table}

\subsubsection*{\textbf{Variability and Thermal Noise Effects}}

The main goal of this paper is to provide the means towards energy effective spin wave gate cascading and enable the design of  spin wave domain circuits.  In view of this we validated our proposal as a proof of the concept without taking into account the influence of edge roughness, waveguide dimension variations, spin wave strength variation, and thermal noise effect. However,  edge roughness and waveguide trapezoidal cross section effects have been investigated and their small impact demonstrated, as the considered gates continued to correctly function even under their presence \cite{DC,DC9}. Furthermore, the thermal noise effect was investigated \cite{DC}. The simulation results indicated that the thermal noise have limited effect on the gate functionality, and that the gate functions correctly at different temperature. The investigation of variability and thermal noise effects one our proposal constitutes future work, even-though we expect that they will have limited impact on spin wave circuit designs. 

\section{Conclusions}
\label{sec:Conclusion}
In conclusion, we proposed a novel  conversion free SW gate cascading scheme that achieves SW amplitude normalization by means of a directional coupler.  After introducing the normalization concept, we utilized if for the implementation of three simple $2$ cascaded Majority gate circuits and of a $2$-bit inputs SW multiplier. We validated the proposed structures by means Object Oriented Micromagnetic Framework (OOMMF) and GPU-accelerated Micromagnetics (MuMax3) simulations. Furthermore, we assessed the normalization induced energy overhead and demonstrated that the proposed approach provides a $20$\% to $33$\%  energy  reduction when compared with the transducers based conventional gate cascading counterpart. Finally, we introduced a normalization based SW $2$-bit inputs multiplier design and compare it with functionally equivalent state-of-the-art designs. Our evaluation indicated that the proposed scheme provided $26$\% and $6.25$x energy reductions when compared with transducers based and  \SI{16}{nm} CMOS counterpart, respectively, which demonstrated the energy effectiveness  of our proposal and its significant contribution towards the full utilization of the SW paradigm potential and the development of SW only circuits.

\begin{acknowledgments}

This work has been funded by the European Union’s Horizon 2020 research and innovation program within the FET-OPEN project CHIRON under grant agreement No.~801055.

\end{acknowledgments}

\bibliography{Spin_Wave_Normalization_Towards_all_Magnonic_Circuits}

\end{document}